\documentclass[pre,aps,onecolumn,amsmath,amssymb,letterpaper,12pt]{revtex4}

\pdfoutput=1
\usepackage{hyperref}
\usepackage{graphicx}

\begin{document}
\title{Approximating the dynamical evolution of systems of strongly-interacting overdamped particles}

\author{Stephen Whitelam\email{swhitelam@lbl.gov}}
\affiliation{The Molecular Foundry, Lawrence Berkeley National Laboratory, Berkeley, CA 94720, USA}

\begin{abstract}
We describe collective-move Monte Carlo algorithms designed to approximate the overdamped dynamics of self-assembling nanoscale components equipped with strong, short-ranged and anisotropic interactions. Conventional Monte Carlo simulations comprise sequential moves of single particles, proposed and accepted so as to satisfy detailed balance. Under certain circumstances such simulations provide an approximation of overdamped dynamics, but the accuracy of this approximation can be poor if e.g. particle-particle interactions vary strongly with distance or angle. The twin requirements of simulation efficiency (trial moves of appreciable scale are needed to ensure reasonable sampling) and dynamical fidelity (true in the limit of vanishingly small trial moves) then become irreconcilable. As a result, single-particle moves can underrepresent important collective modes of relaxation, such as self-diffusion of particle clusters. However, one way of using Monte Carlo simulation to mimic real collective modes of motion, retaining the ability to make trial moves of reasonable scale, is to make explicit moves of collections of particles. We will outline ways of doing so by iteratively linking particles to their environment. Linking criteria can be static, conditioned upon properties of the current state of a system, or dynamic, conditioned upon energy changes resulting from trial virtual moves of particles. We argue that the latter protocol is better-suited to approximating real dynamics.

\end{abstract}

\maketitle

\def\cal#1{\mathcal{#1}}
\def\av#1{\langle #1 \rangle}
\def\b#1{\boldsymbol{#1}}
\def\eqq#1{Eq.~(\ref{#1})}
\def\eq#1{(\ref{#1})}
\def\px{\phi(\b{x})}
\def\tt{|\b{\tau}|}

\def\k{{\kappa}}
\newcommand{\e}{\epsilon}

\def\s{S}
\def\tmc{\tau_{\rm MC}}
\def\rh{\hat{r}}
\def\kt{k_{\rm B} T}
\def\kb{k_{\rm B}}

\def\bef{\beta_{\rm f}}

\def\de{\partial_e}
\def\d{\,\textrm{d}}
\def\es{\boldsymbol{S}}
\def\t{\boldsymbol{\tau}}
\def\r{\boldsymbol{r}}
\def\rr{\hat{\boldsymbol{r}}_{ij}}
\def\v{\boldsymbol{v}}
\def\al {\alpha}
\def\inte{\int\!\!\!\d\e\,}
\def\l {\lambda}
\def\lama {\lam_{\al}}
\def\lamb{\lam_{\beta}}
\def\D {\Delta}
\def\l{\lambda}
\def\L {\Lambda}
\def\on{^{(n)}}
\newcommand{\p}{^\prime}
\def\del{\partial}

\newcommand{\en}{\epsilon_{\rm n}}
\newcommand{\ed}{\epsilon_{\rm d}}

\def\ha{\frac{1}{2}}
\def\beq{\begin{equation}}
\def\eeq{\end{equation}}
\def\bea{\begin{eqnarray}}
\def\eea{\end{eqnarray}}

\def\kt{k_{\rm B}T}

\section{Introduction}

Pairwise-interacting components in implicit solvent are often used as models of self-assembling systems, such as crystal-forming~\cite{wolde1997epc,doye2007controlling,liu2009self} and capsid-forming-~\cite{hagan2006dynamic,nguyen2007deciphering} proteins, and patchy nanoparticles~\cite{patchy2,sciortino2009phase}. In this paper we shall summarize ways of using Monte Carlo (MC) simulation to evolve such components in order to approximate the (interacting) Brownian motion that their real counterparts execute. Brownian motion is usually approximated in simulations by integration of overdamped equations of motion, called the Brownian Dynamics (BD) method~\cite{frenkel2002understanding,rapaport2004art}, with Monte Carlo algorithms more often used as a means of sampling thermal distributions. However, recent work shows that in certain circumstances the MC method can also evolve components according to an approximately correct dynamics. Dynamic MC simulations even offer some advantages over their BD counterparts: it is easier and computationally cheaper to evaluate potentials (MC) than forces (BD); MC can cope with pathological potentials (e.g. hard particles, abrupt changes in potential); and one does not face problems of numerical instability with MC as one does with BD (even for smooth potentials), and so can make larger basic moves.

In what follows we outline the reasons why single-particle Metropolis MC can effect a dynamics close to Brownian motion, and we summarize the work of others that makes use of this correspondence. We then describe extensions of the MC scheme that incorporate explicit moves of collections of particles, and argue that such schemes can be used to preserve the approximate realism of the MC method when collective modes of motion become important to a system's evolution. We take the view that although MC methods are not dynamically realistic in all details, they offer for some applications a convenient alternative to conventional integration of equations of motion.

\section{The dynamical character of Monte Carlo motion}
It is well-known that sequential moves of single components, proposed in an unbiased fashion and accepted according to the Metropolis criterion, permits sampling of the Boltzmann distribution given sufficiently long simulation times~\cite{frenkel2002understanding}. However, is also true that if we restrict such moves to local translations and rotations then the dynamics executed by a single particle in an external forcefield is equivalent, in the limit of small trial moves, to a Langevin dynamics~\cite{mc_dynamics1,mc_dynamics2}, i.e. to Brownian motion in that potential. To see this, consider a particle in one dimension in a position-dependent potential $U(x)$. The master equation corresponding to a Metropolis MC algorithm in which particle displacements $\hat{x} \equiv x'-x$ are drawn uniformly from a range $[-\Delta,\Delta]$ is
\beq
\label{master}
\partial_t P(x;t)= \int_{-\Delta}^{\Delta} {\rm d}\hat{x} P(x';t) W(x'\to x) -\int_{-\Delta}^{\Delta} {\rm d}\hat{x}  P(x;t) W(x\to x'). 
\eeq
Here $P(x;t)$ is the probability of finding the particle at position $x$ at time $t$, $W(x \to x')=(2 \Delta)^{-1} \min \left(1,\exp\left[-\beta U(x')+ \beta U(x) \right]\right)$ is the rate of moving a particle from position $x$ to position $x'$, and $\beta \equiv 1/(\kt)$. \eqq{master} can be expanded in powers of $\Delta$ (which we assume to be small; we also assume that $U'' \cdot\Delta\ll U'$), giving to lowest order
\beq
\label{lang}
\partial_t P(x;t) \approx -\partial_x \left( v P(x,t) \right)+\partial_x \left(D \partial_x P(x,t) \right).
\eeq
This is a Fokker-Planck equation with drift velocity $v =- (\beta/6)   \Delta^2 \, U'(x) +{\cal O}(\Delta^3)$ and diffusion constant $D = \Delta^2/6+{\cal O}(\Delta^3)$, and corresponds to a Langevin dynamics satisfying an Einstein relation $-v/D \approx \beta \, U'(r)$. We can neglect terms higher order in $\Delta$ provided that $U(x)$ changes little in the course of a single move. If this condition holds then Metropolis MC moves of single particles in an external potential occur in a dynamically realistic way: the drift velocity of the particle is proportional to the force acting upon it, and its diffusion constant is independent of position. This correspondence also holds in two and three dimensions.

In most situations we are interested in interacting particles, not isolated particles in external potentials. Here, too, single-particle Metropolis MC evolution can in many cases approximate a realistic dynamics. As an illustration, consider two interacting but otherwise isolated particles $i$ and $j$ in one dimension, with positions $x_i$ and $x_j$. Particles interact according to a pairwise potential $U(x_i-x_j)$. Under the single-particle Metropolis Monte Carlo algorithm described before, the master equation for the separation $r\equiv x_i-x_j$ of these particles reads
\beq
\label{ma2}
\partial_t P(r;t)= \int_{-\Delta}^{\Delta} {\rm d}\hat{r} P(r';t) W(r'\to r) -\int_{-\Delta}^{\Delta} {\rm d}\hat{r}P(r;t) W(r\to r'),
\eeq
where $\hat{r} \equiv r'-r$ and $W(r \to r')=(2 \Delta)^{-1} \min \left(1,\exp\left[-\beta U(r')+ \beta U(r) \right]\right)$. As before, expansion of this equation in powers of $\Delta$ yields a Fokker-Planck equation with drift velocity $v \approx -\Delta^2 \beta \,U'(r)/6$ and diffusion constant $D \approx \Delta^2/6$. These results are proportional to those obtained by assuming that $i$ and $j$ are subject to a Brownian motion described by the equations
\beq
\label{eq1}
\gamma \dot{x}_i = -\partial_{x_i} U(r) +\eta_i 
\eeq
and
\beq
\label{eq2}
\gamma \dot{x}_j = -\partial_{x_j} U(r) +\eta_j,
\eeq
where the noise terms $\eta_{i,j}$ have zero mean and variance $\langle \eta_i(t) \eta_j(t') \rangle = 2 k_{\rm B}T \gamma \delta_{ij} \delta(t-t')$, and $\gamma$ is a friction coefficient. Further, the master equation for the center of mass $R \equiv \frac{1}{2} (x_i+x_j)$ of the dimer $ij$ reads
\bea
\label{dimer}
\partial_t P(R;t)&=& \int_{-\Delta}^{\Delta} {\rm d}\hat{r} P(R-\hat{r}/2;t) \left[ W(x_i -\hat{r}\to x_i,x_j)+ W(x_i,x_j-\hat{r} \to x_j)\right] \nonumber \\
 &-&\int_{-\Delta}^{\Delta} {\rm d}\hat{r}P(R;t) \left[ W(x_i \to x_i+\hat{r},x_j) +W(x_i, x_j \to x_j+\hat{r}) \right],
\eea
where $W(x_i -\hat{r}\to x_i,x_j)=(4 \Delta)^{-1} \min \left(1,\exp\left[-\beta U(x_i-x_j)+ \beta U(x_i-\hat{r}-x_j) \right]\right)$. Expansion of \eqq{dimer} yields $\langle R \rangle =0$ (since no external forces act on the dimer) and $\langle R^2 \rangle =\Delta^2/(6 \cdot 4) + {\cal O}(\Delta^3 U'(r))$. Hence in the limit of small displacements $\Delta$ the collective diffusion constant is independent of the force exterted by $i$ on $j$, which is what one would conclude by adding Eqs.~\eq{eq1} and~\eq{eq2}. If, by contrast, trial moves of $i$ and $j$ lead to large energy changes, then the dimer diffusion constant $\langle R^2 \rangle \propto \int_{-\Delta}^{\Delta} {\rm d}\hat{r} \,\hat{r}^2 \exp\left[-\beta U(r')+ \beta U(r) \right]$
 will be suppressed (potentially strongly so) relative to the overdamped ideal.

\begin{center}
\begin{figure*}
\includegraphics[width=\linewidth]{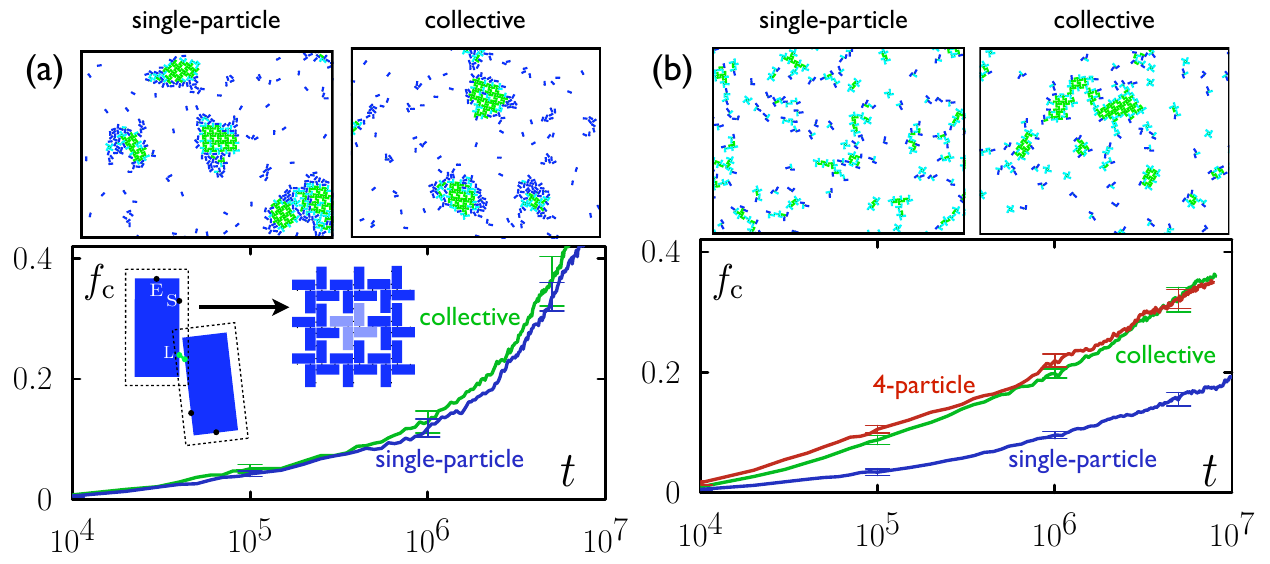} 
\caption{\label{fig1} When explicit collective motion does not (a) and does (b) matter. We show self-assembly trajectories for the model protein system introduced in Ref.~\cite{PhysRevLett.105.088102}, which comprises hard rectangles on a two-dimensional substrate (see inset to (a)). Rectangles attract via a rectangular nonspecific forcefield (which encourages fluidlike clustering of monomers) and via directional linkers mediated by sticky patches labeled E, S and L (which allow the formation of a square lattice of tetramers). Only E-S and L-L pairings are reactive. Here we evolve 1800 monomers at 10\% area fraction using one of three protocols. These are the virtual-move algorithm described in section~\ref{sec_virtual} (`collective'), with basic displacement scale just larger than the maximum range of particle-particle attractions; and the same algorithm in which we reject explicit motion of any cluster larger than a monomer (`single-particle') or a tetramer (`4-particle'). The lines record the fraction of crystalline particles $f_{\rm c}$ (i.e. those with three engaged directional bonds) as a function of the number of simulation timesteps $t$; data are averaged over 8 independent simulations. In (a), the nonspecific attraction is $2.25 \, \kt$ and E-S and L-L bonds confer $4 \, \kt$ of attractive energy. The resulting dynamics comprises transient, precritical liquidlike fluctuations that develop crystalline order within them. While large structures do move collectively if permitted to do so, crystallization is dominated by detachment and attachment of single particles from and to fluidlike blobs, and by rotations of single particles within these blobs. As a result, we see little difference between collective- and single-particle algorithms: the averages of the respective trajectories lie closer than the trajectories' standard deviations. Snapshots, which show about 25\% of the simulation box, are taken at $5 \times 10^6$ timesteps; crystalline particles are green. A qualitatively different parameter regime is shown in panel (b). Here there is no nonspecific attraction; E-S bonds have a strength of $100\,  \kt$; and L-L bonds have a strength of $8.25 \, \kt$. The strong E-S bonds encourage the rapid formation of tetramers and fragments thereof, and these clusters subsequently assemble into higher-order structures via L-L bonds. We see a large difference between a collective algorithm and a single-particle algorithm, because the latter underrepresents tetramer diffusion. Restoring explicit collective moves of only 4 particles gives rise to behavior similar to that of the fully collective algorithm. Snapshots are taken at $7.5 \times 10^6$ timesteps.}
\end{figure*}
\end{center}

These arguments are difficult to extend analytically to larger numbers of particles, but they suggest that single-particle Metropolis MC can mimic a realistic dynamics A) if trial moves can be made small enough to render negligible higher-order corrections to particles' drift and diffusion coefficients, or B) if collective diffusive modes of motion do not dominate a system's behavior. The latter situation might arise if a system is very crowded, and particles are able to move only small distances, or if assembly of a structure is dominated by the addition and detachment of single particles, rather than by cluster-cluster interactions. We illustrate this second scenario in Fig.~\ref{fig1}(a).

As suggested by these simple arguments, Metropolis Monte Carlo simulations have been shown in several cases to approximate a realistic dynamics. Ref.~\cite{mc_dynamics2} reports good agreement between the dynamics of a model protein evolved by integration of Langevin equations, and by Metropolis MC using small trial moves. Ref.~\cite{sanz2010dynamic} demonstrates near-quantitative agreement between Brownian dynamics and Metropolis MC algorithms used to calculate the self-diffusion constant of a colloidal fluid and its crystallization dynamics. Metropolis simulations of the crystallization of attractive colloids, in parameter regimes dominated by single-particle attachment and detachment, were found to behave like their Brownian counterparts~\cite{scarlett2010computational}. Refs.~\cite{mc_dynamics3,mc_dynamics4} even found accord between the long-time equilibrium dynamics of silica computed with Metropolis MC and with Newtonian (not overdamped) dynamics, perhaps because the distinction between diffusive and ballistic dynamics is relatively unimportant in a crowded system. Refinement of the dynamical realism of single-particle MC trajectories can be achieved by rescaling the basic MC timestep by acceptance rate~\cite{heyes1998monte, sanz2010dynamic}, or by biasing the choice of trial moves according to the forces acting on particles~\cite{rossky1978brownian}.
  
\section{Collective moves: static linking schemes}

There also exist situations in which collective diffusive modes of motion are important to a system's evolution, and it is inconvenient to make the Monte Carlo trial step size small enough that a single-particle algorithm can mimic realistic collective diffusion. Such situations can arise if particles' interactions change rapidly with distance or angle (true of e.g. certain model proteins, which might extend several nanometers but interact over distances comparable to a nanometer); if particles are present at low concentration (because they must cross large distances to interact); and if a system's assembly is naturally hierarchical. We show in Fig.~\ref{fig1}(b) an example in which all three of these conditions holds. 

When faced with such a system, one can restore to a Monte Carlo scheme a degree of realistic collective motion by making explicit moves of collections of particles. This is often done by recursively linking particle pairs (starting, say, with particles $i$ and $j$) with a given probability, and proposing a collective move of the resulting cluster~\cite{panag_cluster,Liu1, SW,wu,wolff}. This procedure connects an initial microstate, $\mu$, with a proposed new one, $\nu$. We shall write $p_{ij}(\mu \to \nu)$ for the probability of linking $i$ and $j$ in the course of proposing the move $\mu \to \nu$. 

\begin{center}
\begin{figure}
\includegraphics[width=0.8\linewidth]{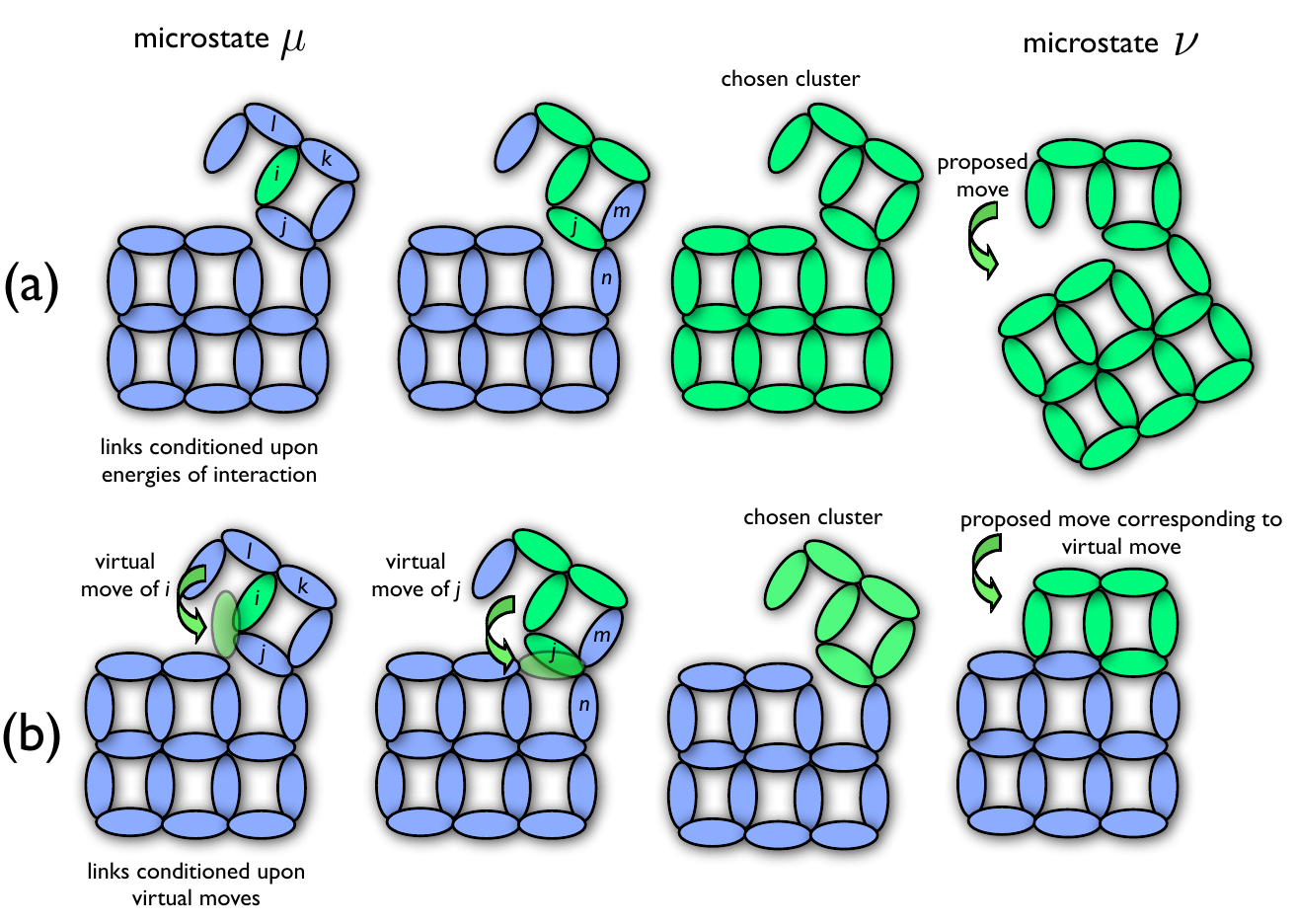} 
\caption{\label{fig2} Illustration of static and dynamic linking procedures for cluster moves. In (a), nanoparticle $i$ is linked to its neighbors $j$, $k$ and $l$ according to pairwise energies of interaction in the initial microstate, $\mu$. All recruited neighbors (e.g. $j$) propose links with their neighbors (e.g. $m,n$), and so on until no particles remain to be tested. In the example shown all particles interact strongly, and the entire cluster is chosen to move. The move shown results in proposed new microstate $\nu$. In (b), $i$ is recursively linked to its environment according to gradients of interaction energies calculated by making virtual moves of particles (see text). The proposed move leads to new microstate $\nu$.}
\end{figure}
\end{center}

As Fig.~\ref{fig2} illustrates, particles can be linked statically or dynamically. In a static linking scheme (see e.g.~\cite{SW,wu, whitelam2007avoiding,bhattacharyay2008self}) the likelihood of linking $i$ and $j$ depends only on their properties (e.g. energy or degree of proximity) in the original microstate $\mu$, i.e. $p_{ij}(\mu \to \nu) = p_{ij}(\mu)$. We sketch one example of this procedure in Fig.~\ref{fig2}(a). We shall specialize our discussion in this section to the simple case of a pairwise potential $\epsilon_{ij}$ composed of a hard-core part plus an attractive part; one possible static linking scheme is as follows. We choose an initial seed particle $i$, and attempt to link this particle to one of its neighbors $j$ with probability
\beq
\label{link}
p_{ij}(\mu \to \nu)=\Theta \left(n_{\rm c} -n_{{\cal C}}  \right) \cal{I}_{ij}(\mu)\left(1-{\rm e}^{\bef \epsilon_{ij}}\right),
\eeq
where $\bef$ is a fictitious reciprocal temperature. The factor $\cal{I}_{ij}(\mu)$ is unity if $i$ and $j$ interact (e.g. lie within possible interaction range) in microstate $\mu$, and zero otherwise; the factor $\Theta \left(n_{\rm c} -n_{{\cal C}}  \right)$ terminates link formation if the number of particles recruited to the cluster, $n_{{\cal C}}$, exceeds $n_{{\rm c}}$, the smallest integer larger than $\xi ^{-1}$. Here $\xi $ is a random variable drawn uniformly from the interval $[0,1]$. If $n_{{\cal C}}$ exceeds $n_{{\rm c}}$ then the move is aborted {\em in situ}, preventing clusters from moving with a frequency greater than is physical.

If the link is succesful then we add $j$ to the moving cluster; if not, we do not, and we do not attempt to link $i$ to $j$ again. We continue iteratively to propose links between particles in the moving cluster and those with which they interact (as long as we have not tested those links before, and provided that those particles are not already members of the moving cluster). We stop when we run out of particles to test. We then propose a move (e.g. a translation) of the cluster. This defines a new microstate $\nu$. To preserve the equilibrium distribution it is sufficient to impose the requirement of superdetailed balance~\cite{super_duper},
\beq
\label{superdetbal}
\rho(\mu) W_{\rm gen}(\mu \to \nu|{\cal R}) W_{\rm acc}(\mu \to \nu|{\cal R}) = \rho(\nu) W_{\rm gen}(\nu \to \mu| \cal{R}) W_{\rm acc}(\nu \to \mu| \cal{R}).
\eeq
Here $\rho(\mu)={\rm e}^{-\beta E(\mu)}/Z$ is the equilibrium weight of the state $\mu$ ($E(\mu)$ is the energy of state $\mu$ and $Z$ is the partition function), and $W_{\rm gen}(\mu \to \nu |{\cal R})$ is the rate of generating a move from state $\mu$ to state $\nu$, given a realization $\cal{R}$ of links and failed links. This rate contains the likelihood of selecting the cluster's displacement or rotation, one factor of $p_{ij}(\mu)$ for each link formed within the moving cluster, one factor of $1-p_{ij}(\mu)$ for each link attempted but not formed within the cluster, and one factor of  $1-p_{ij}(\mu)$ for each link not formed between the cluster and its environment. All but the latter set of probabilities equal their counterparts for the reverse move. Rearranging \eqq{superdetbal} reveals that balance is satisfied by the acceptance rate 
 \beq
\label{accept1}
W_{\rm acc}(\mu \to \nu)= {\cal D}({\cal C})\min \left(1, {\rm e}^{\left(\bef-\beta\right) \left( E(\nu)-E(\mu) \right)}\right),
\eeq 
provided that no overlaps occur. If they do, the move is rejected. For infinite fictitious temperature, $\bef =0$, the likelihood of forming links between the seed $i$ and any other particle is zero, and the algorithm executes single-particle moves. For finite values of $\bef$, collective moves can be achieved. We have attached a factor of $\cal{D}({\cal C}) \leq 1$ in order to modulate the diffusivity of clusters according to their size and shape.

Such schemes provide a convenient way to effect collective motion in self-assembling systems~\cite{jack2007fluctuation,bhattacharyay2008self,jankowski2009comparison,liu2009self,mondal2010sequence}. They allow for precise control of collective motion: we know in advance of the move the nature of the cluster to be moved, and we can rotate and translate this cluster as desired. Their chief shortcoming, however, is that clusters (and single particles) do not move solely according to the potential energy gradients acting on them. Because links are conditioned upon energies in the initial microstate, particles interacting strongly are likely to be moved in concert, even if relative moves of these particles are favorable. The analog of \eqq{dimer} for a static cluster algorithm yields an effective drift velocity for the inter-dimer separation that is not simply proportional to the negative of the potential gradient, but is instead proportional to  $-\epsilon_{ij}'(r)  {\rm e}^{\bef \epsilon_{ij}(r)}$. This drift velocity is not consistent with a physical dynamics. One suggested consequence of such a bias is sketched in Fig~\ref{fig2}(a): even though rotation of the upper 6 particles might be desirable (see panel (b)), if particles interact strongly then a static algorithm would have trouble forming a cluster of those 6 particles that does not include the whole of the structure shown. Simulations~\cite{martsinovich2010modeling} show that proposing relative moves of strongly-attracting particles less frequently than moves of weakly-attracting particles can lead to dynamical trajectories substantially different than are generated by integrating equations of motion.

\section{Collective moves: dynamic linking schemes}
\label{sec_virtual}

By contrast, the idea behind a dynamic cluster-linking scheme~\cite{wolff} is to make a trial move of a single particle and to deal iteratively with the consequences of that move. This scheme, and certain of its off-lattice generalizations~\cite{Liu1,whitelam2007avoiding}, decouple the likelihood of proposing relative moves of particles from their interaction energies in the initial microstate, circumventing the chief deficiency of static cluster-linking schemes. In Fig.~\ref{fig2}(b) we illustrate one possible dynamic cluster-linking scheme~\cite{whitelam2009role}, called a `virtual-move' Monte Carlo algorithm, for particles bearing general pairwise interactions $U_{ij}$. We link particles in a recursive manner similar to that described above, except that now our linking procedure involves trial virtual moves of particles. In detail, we pick a particle $i$. We choose to form a pre-link of particle $i$ and some neighbor $j$ with a probability
\beq
\label{virtual_link}
p_{ij}(\mu \to \nu)=\Theta \left(n_{\rm c} -n_{{\cal C}}  \right) \cal{I}_{ij}(\mu) \textnormal{max}\left(0,1-{\rm e}^{\beta U_{ij}(\mu)-\beta U_{i'j}(\mu)}\right)
\eeq
that depends on a {\em virtual move} (e.g. translation or rotation) of $i$ relative to $j$. Here $U_{ij}(\mu)$ is the pairwise energy of the bond $ij$ in microstate $\mu$, and  $U_{i'j}(\mu)$ is the bond energy following the virtual move of $i$ ($i$ is returned to its original position following its virtual move). The factors $\cal{I}_{ij}(\mu)$ and $\Theta \left(n_{\rm c} -n_{{\cal C}}  \right)$ are as before, and as before the move is aborted if $n_{{\cal C}}$ exceeds $n_{{\rm c}}$. Linking particles in this fashion ensures that neighbors exert mutual forces proportional to the gradient of their pairwise energies, unlike in static linking schemes. Particle motions are correlated at the level of a single move if $\beta U_{i'j}(\mu) -\beta U_{ij}(\mu)$ is large. 

We then do as follows.

\begin{itemize}
\item If a pre-link does not form, we label the link $ij$ as unformed, do not add $j$ to the moving cluster, and do not consider the link $ij$ again. We then consider another neighbor of $i$. 
\item If the pre-link forms, then
\begin{itemize}
\item we convert the pre-link into a full link with probability $f_{\rm reverse}(\mu \to \nu) \equiv \min \left(1, \frac{p_{ij}(\nu \to \mu) }{p_{ij}(\mu \to \nu)} \right)$, and add $j$ to the moving cluster. $j$ is then assigned a virtual move so that it moves with $i$ as a rigid body. To compute $f_{\rm reverse}$ we make a reverse virtual move of $i$ (starting from its original position), corresponding to the forward virtual move with the sense of rotation or translation reversed. For pre-linked particles (indeed, for any two particles internal to the chosen cluster) the factor $p_{ij}(\nu \to \mu)$ is given by \eqq{virtual_link} where $i'$ now refers to $i$ following a reverse virtual move. 
\item we convert the pre-link into a frustrated link with probability $1-f_{\rm reverse}(\mu \to \nu)$. In this case $j$ is not added to the moving cluster, and the bond $ij$ is not tested again. 
\end{itemize}
\end{itemize}
We stop when no more particles remain to be tested, and we move the chosen cluster according to the prescribed virtual move, defining a proposed new microstate $\nu$.

To preserve the equilibrium distribution we can balance the rates for forward and reverse moves involving a given realization ${\cal R}$ of 1) internal cluster links, and 2) failed internal links that are either unformed or frustrated. By construction of the linking procedure these two classes of probabilities cancel from \eqq{superdetbal}. The remaining contribution to \eqq{superdetbal} comes from unformed links external to the moving cluster, and rearrangement of that equation reveals that an appropriate acceptance rate for the collective move is
\bea
\label{accept2}
W_{\rm acc}{(\mu \to \nu|\cal{R})} =  {\cal D}(\cal{C}) \min \left(1,  \prod_{\langle i j \rangle_{{\rm n}\leftrightarrow {\rm o}}} {\rm e}^{-\beta\left( U_{ij}(\nu) -U_{ij}(\mu) \right)} \right),
\eea
provided that no frustrated links lie external to the pseudocluster; the acceptance rate is zero if they do. The label $\langle ij \rangle_{{\rm n} \leftrightarrow {\rm o}}$ identifies particle pairs that start ($\mu$) in a noninteracting configuration and end ($\nu$) with positive energy of interaction (overlapping), {\em or} which start ($\mu$) in an overlapping configuration and end ($\nu$) in a noninteracting one.

There is no guarantee that such a procedure will result in motion that is dynamically realistic in all details. Indeed, there are some features of the algorithm that make precise control of cluster motion impossible. For one, we do not know in advance the nature of the moving cluster, making it hard to cleanly separate translational from rotational motion (the same is not true of a static linking procedure). For another, collective motion requires that both forward and reverse virtual moves of one particle `recruit' another: if the basic scale of virtual displacements is much smaller than particle interaction ranges then such motion is unlikely. Correspondingly, if the basic scale of virtual displacements is too large, intra-cluster relaxation becomes slow. Some tinkering is needed in order to reach a reasonable compromise between the these two processes.

Even given these difficulties, intuition suggests that by moving clusters according to gradients of potential energy, and by choosing cluster diffusion constants $D({\cal C})$ in a reasonable way, we should reproduce some key features of overdamped motion. (It is also worth noting that we have the freedom to scale collective diffusion constants $D({\cal C})$ anisotropically, as is physically reasonable, which is not a feature that emerges from simple BD algorithms.) Particles should move in a locally realistic fashion and retain some of the collective degrees of freedom that single-particle moves ignore. A qualitative comparison between a virtual-move algorithm (the version of Ref.~\cite{whitelam2007avoiding}, which is the predecessor of the algorithm described here) and BD simulations of strongly-attractive discs shows this to be the case, even in circumstances where single-particle moves clearly lack dynamical accuracy~\cite{whitelam2007avoiding}. Preservation of these important features of real dynamics may be sufficient to determine if the real counterpart of a model system will assemble well or become kinetically trapped. Testing of a virtual-move algorithm against BD simulations of viral capsid self-assembly found that each generates similar values of capsid yields for given model parameters~\cite{whitelam2009role}. Since yields depend upon both thermodynamics and dynamics, such agreement is encouraging. Other work has used virtual-move algorithms (the original version~\cite{whitelam2007avoiding} or the one described here) to generate dynamical trajectories for self-assembling systems~\cite{whitelam2009impact,ouldridge2010dna, whitelam2010self,PhysRevLett.105.088102, soliscontrolled1,soliscontrolled2,villar2009self} or to thermodynamically sample them~\cite{russo2010association}.

The cluster schemes described above treat pairwise-interacting particles. However, it is possible to use them to effect collective motion of particles bearing multibody potentials~\cite{rogers2010prep}, which are often encountered in model biomolecules~\cite{vitalis2009methods}. One way to do so is described in~\cite{frenkel2002understanding}. Let's say that the true energy of a system of particles in microstate $\mu$ is $E(\mu)$, which may contain contributions from multibody potentials. We can nonetheless use the virtual-move scheme by assuming that all particles interact via fictitious pairwise potentials $U_{ij}$ (perhaps derived from potentials of mean force obtained using particles' true interactions). If we write $E(\mu)= (E(\mu)-U(\mu))+U(\mu)$ in the exponentials in the equilibrium weights in~\eqq{superdetbal}, where $U(\mu) \equiv \frac{1}{2} \sum_{ij} U_{ij}(\mu)$ is the system's total fictitious energy in microstate $\mu$, we find the acceptance rate for the virtual-move procedure using the fictitious potentials $U_{ij}$ to be
\bea
\label{accept3}
W_{\rm acc}{(\mu \to \nu|\cal{R})} =  {\cal D}(\cal{C}) \min \left(1,  {\rm e}^{-\beta\left(\Delta E-\Delta U\right)}\prod_{\langle i j \rangle_{{\rm n}\leftrightarrow {\rm o}}} {\rm e}^{-\beta\left( U_{ij}(\nu) -U_{ij}(\mu) \right)} \right),
\eea
subject to the same caveats as~\eqq{accept2}. Here $\Delta E \equiv E(\nu)-E(\mu)$ and $\Delta U \equiv U(\nu)-U(\mu)$. The factor $ {\rm e}^{-\beta\left(\Delta E-\Delta U\right)}$ accounts for the difference between real and fictitious potentials. A fictitious linking potential can also be used if the real potential $E$ contains long range interactions (e.g. $1/r$ Coulomb interactions) that make direct application of a cluster algorithm inconvenient. In this case, the fictitious potential $U$ could be chosen to account only for the short range component of particles' interactions.

\section{Conclusions}

We have described the use of Monte Carlo algorithms to approximate the overdamped dynamics of interacting particle systems. While neither single-particle- nor collective-move algorithms are dynamically realistic in all details, recent work shows that they can approximate a natural dynamics within a range of model systems. Given the advantages of numerical stability and ease of implementation offered by Monte Carlo algorithms over Brownian Dynamics schemes, we suggest that Monte Carlo algorithms can in some cases provide a useful and convenient alternative to conventional integration of equations of motion.

\section{Acknowledgements}

I thank Rob Jack and Jocelyn Rodgers for comments on the manuscript. I am grateful to Phill Geissler for the collaboration that led to development of the virtual-move algorithm (Ref.~\cite{whitelam2007avoiding}), and I thank Alex Wilber, Tom Ouldridge and Jon Doye for identifying omissions in preprint- and published versions of that paper. This work was performed at the Molecular Foundry, Lawrence Berkeley National Laboratory, and was supported by the Director, Office of Science, Office of Basic Energy Sciences, of the U.S. Department of Energy under Contract No. DE-AC02--05CH11231.

\end{document}